\documentclass[prd,preprint,preprintnumbers,showpacs,nofootinbib,amssymb]{revtex4}
\usepackage{graphicx}% Include figure files
\usepackage{dcolumn}% Align table columns on decimal point
\usepackage{bm}% bold math
\usepackage{color}
\usepackage{subfigure}
\newcommand\beq{\begin{equation}}
\newcommand\eeq{\end{equation}}
\newcommand\bea{\begin{eqnarray}}
\newcommand\eea{\end{eqnarray}}
\newcommand\bi{\begin{itemize}}
\newcommand\ei{\end{itemize}}
\newcommand\ben{\begin{enumerate}}
\newcommand\een{\end{enumerate}}

\newif\ifboo \boofalse

\def\lsim{\mathrel{\rlap{\lower4pt\hbox{\hskip1pt$\sim$}}
    \raise1pt\hbox{$<$}}}         %less than or approx. symbol
\def\gsim{\mathrel{\rlap{\lower4pt\hbox{\hskip1pt$\sim$}}
    \raise1pt\hbox{$>$}}}         %greater than or approx. symbol
\begin{document}
\begin{flushleft} 
TTK-11-31 \newline
RECAPP-HRI-2011-005
\end{flushleft}
\title{\Large{Constraints on axino warm dark matter from X-ray
    observation at the Chandra telescope and SPI}}
\author{\bf Paramita Dey} 
\email{paramita@physik.rwth-aachen.de}
\affiliation{Institut f$\ddot{u}$r
  Theoretische Teilchenphysik und Kosmologie, \\
  RWTH Aachen, D-52056 Aachen, Germany} 
\author{\bf Biswarup
  Mukhopadhyaya} 
\email{biswarup@hri.res.in}
\affiliation{Regional Centre for Accelerator-based Particle Physics, \\
  Harish-Chandra Research Institute, Chhatnag Road, Jhusi, Allahabad
  211019, India} 
\author{\bf Sourov Roy}
\email{tpsr@iacs.res.in} 
\affiliation{Department of Theoretical Physics,
  Indian Association for the Cultivation of Science, \\ 2A $\&$ 2B Raja
  S.C. Mullick Road, Kolkata 700032, India} 
\author{\bf Sudhir
  K. Vempati} 
\email{vempati@cts.iisc.ernet.in} 
\affiliation{Centre
  for High Energy Physics, Indian Institute of Science, Bangalore 560012, India}
\date{\today}
\begin{abstract}
  A sufficiently long lived warm dark matter could be a source of
  X-rays observed by satellite based X-ray telescopes. 
  We consider axinos and gravitinos with masses between 1 keV and 100 keV in
  supersymmetric models with small R-parity violation.  We show that
  axino dark matter receives significant constraints from X-ray
  observations of Chandra and SPI, especially for the lower end of the allowed range of
  the axino decay constant $f_{a}$, while the gravitino dark matter
  remains unconstrained.
\end{abstract}


\maketitle
%\vskip .6 true cm
%------------------------------------------------------------------------
\section{Introduction}
Though many popular supersymmetric (SUSY) models tout the lightest
neutralino as the dark matter candidate, the axino or the gravitino,
too, may qualify for a similar role.  The stability of these dark matter
candidates is generally ensured by the conservation
of $R$-parity, defined by $R = (-1)^{(3B+L+2S)}$, where
$B$, $L$ and $S$ are baryon number, lepton number and spin
respectively. $R$-parity keeps the proton from
decaying. As a by-product, it keeps the lightest supersymmetric
particle (LSP) stable. 

This, however, need not always be the case. It is sufficient for the
dark matter particle to be stable on cosmological scales rather than
be fundamentally stable. Thus a slowly decaying light axino or
gravitino in an R-parity violating theory is not incompatible with a
universe containing dark matter.  Such a situation has a
distinct phenomenology of its own.

Since the simultaneous violation of B and L would lead to unacceptably
fast proton decay, one usually assumes only one of them to be violated
by odd units. L-violation, in particular, leads to useful mechanisms
for the generation of neutrino Majorana masses at both the tree- and
one-loop levels. 

{\it Prima facie},  L-violation by odd units 
renders the LSP unstable. This rules out the candidature
of the lightest neutralino as a dark matter constituent as
it tends to decay fast compared to the cosmological scales.  On the
other hand, if the LSP is a light gravitino or an axino, it is still possible
to have a viable dark matter candidate. The requirement is that the decay 
of the gravitino or the axino through R-parity violating couplings should be
sufficiently suppressed such that their lifetime is larger than the
age of the universe. In this note, we explore this possibility and
study the consequent astrophysical observations, in terms of
X-rays coming from galaxies and galaxy-clusters.

The axion field in a supersymmetric model is the pseudoscalar part of
an electrically neutral chiral supermultiplet $\Phi_a$. $\Phi_a$ can
be expanded as $\Phi_a = (s +ia)/\sqrt{2} + \theta {\tilde a} + \theta
\theta F$, where the scalar field is denoted by $s$, the axion, or the
pseudoscalar field, by $a$ and their fermionic partner, the axino, is
denoted by ${\tilde a}$. As long as SUSY is unbroken, the supermultiplet 
remains light, since it is protected by the Peccei-Quinn synmetry $U(1)_{\rm PQ}$ 
\cite{kim-masiero-nanopoulos,chun-kim-nilles,chun-lukas}. The masses 
split, lifting the axino  from the almost massless axion, if
SUSY is broken. This is achieved either at the tree level or via loop
diagrams. The precise value of the axino mass depends on the
model. While one would typically expect the axino mass to be around
$M_{\rm SUSY}$, the soft SUSY breaking scale,  it has been realised
\cite{frere-gerard,kim-masiero-nanopoulos,chun-kim-nilles} that it can
be much smaller than $M_{\rm SUSY}$ depending on the mechanism of mass
generation. In the present work, we set the axino mass as a free
parameter, which could span many classes of models.

It is interesting to note that the axino is a good dark matter
candidate even if R-parity is not exactly conserved. The lifetime of
the axino can be very long because of the suppression from a large
Peccei-Quinn (PQ) symmetry breaking scale and a small amount of
R-parity violation. Axino cold dark matter (CDM) with R-parity
breaking has been considered in \cite{kim-kim} and subsequently in the
context of the INTEGRAL anomaly in \cite{Integral}.  For non-thermal
production, the axinos with a mass in the range 2--100 keV can be the
warm dark matter (WDM)
\cite{axino-warm-dark-matter,seto-yamaguchi-axino-warm-dm}.

Another candidate for warm dark matter in supersymmetric theories is
the gravitino.  Although the mass of the gravitino in the simple types
of supergravity (mSUGRA) models is in the electroweak/TeV scale,
situations with a very light gravitino, too, are often envisioned in
supersymmetry. For example, light gravitinos are natural in models
with gauge mediated supersymmetry breaking. Such a light gravitino can
be the dark matter candidate under certain circumstances
\cite{gravitinoref}.  These light gravitinos can lead to novel
signatures \cite{gravitinoref,biswas-chakrabortty-roy} at colliders
and accelerators and further have strong astrophysical and
cosmological implications \cite{gravitinoref}.

The light LSP's we are considering \textit{i.e.}  the gravitino/axino
could form either cold dark matter or warm dark matter depending on
their mass ranges. Typically a $\sim $GeV dark matter would be a CDM
whereas a keV dark matter would be a WDM.  The exact limits however
depend on the details of the candidate, its couplings etc.  An axino
with masses between $2$ keV to $100$ keV has been shown to be
consistent as a warm dark matter candidate satisfying the relic
density constraints \cite{axino-warm-dark-matter}.  In the present
work we will consider this mass range for the axino, however, with
R-parity violation.

A particularly simple and minimal model of R-parity violation is the
case where there are \textit{only} bilinear $\Delta L =1$ terms in the
superpotential,
\begin{equation}
W_{\not{L}} = \epsilon_i L_i H_u,
\end{equation}
where $L_i$ and $H_u$ are the leptonic and up-type Higgs super-fields
and the index $i=\{1,2,3\}$ runs over the generations. The model which
has been studied over the years from many angles \cite{brpv} is
specified by six additional parameters at the weak scale: the three
sneutrino vacuum expectation values (VEV) and three bilinear R-parity
violating couplings. The presence of these bilinear terms leads the
axino to decay into a photon and a neutrino with the width given as
\cite{axino-warm-dark-matter}:
\begin{equation}
\label{axino-2-body-decay}
\Gamma( {\tilde a} \to \gamma \nu_i) = {\frac{1}{128 \pi^3}}~ | U_{\tilde{\gamma} {\tilde Z}} |^2 ~ 
{\frac{m_{\tilde a}^3}{f^2_a}} ~ C^2_{a\gamma\gamma} ~ \alpha^2_{em}~ \xi^2_i,
\end{equation}
where $\xi_i = {\langle {\tilde \nu}_i \rangle}/v$ with the Higgs VEV given by $v$ = 174 GeV,
\begin{equation}
U_{{\tilde \gamma} {\tilde Z}} = M_Z \sum_\alpha \frac{N_{{\tilde
      Z}\alpha} N^*_{{\tilde \gamma}\alpha}}{m_{{\tilde \chi}_\alpha}}
\end{equation}
is the photino-zino mixing parameter containing the neutralino mixing
matrix N and the mass eigenvalues $m_{{\tilde \chi}_\alpha}$, $\alpha_{em}$ is
the fine structure constant, $M_Z$ is the mass of the Z-boson, $m_{\tilde a}$ is the
axino mass, $f_a$ is the Peccei-Quinn symmetry breaking scale (up to a domain wall number),
also identified with the axion decay constant, and
$C_{a\gamma\gamma}$ is axion-photon-photon coupling, a model dependent
quantity (See Kim \cite{kimprd} for a table of values). If we restrict
ourselves to hadronic axion models, the axino decay into a photon and
a neutrino arises from the anomaly coupling with the photon vector
multiplet and the neutralino-neutrino mixing generated by the
sneutrino VEV $\langle {\tilde \nu}_i \rangle$
\cite{axino-warm-dark-matter}.

For a suitably small mixing parameter, it is remarkable that the lifetime 
of the axino can exceed that of the age of the universe, for a
Peccei-Quinn symmetry breaking scale of $\mathcal{O}(10^{11}) ~\text{GeV}$ and
an axino mass of 5 keV. This can be seen from the life time expression
for the axino given as:
\begin{equation}
 \tau( {\tilde a} \to \gamma \nu_i) = 3 \times 10^{35}s 
\left(\frac{\xi_i |U_{{\tilde \gamma} {\tilde Z}}|}{10^{-7}}\right)^{-2} 
  \left(\frac{m_{\tilde a}}{5~{\rm keV}}\right)^{-3} 
\left(\frac{f_a}{10^{11}~{\rm GeV}}\right)^2.
\label{axino-life-time}
\end{equation}

In a similar manner, gravitino can also live long enough to be
considered as dark matter though it is governed by a different set of
couplings. In a recent analysis by Buchmueller et. al \cite{ibarra} it
has been shown that gravitinos of mass $\mathcal{O}(5~{\rm GeV})$ or just
above are compatible with thermal leptogenesis, cold dark matter and
Big Bang Nucleosynthesis.  However, purely from a dark matter relic
density point of view, it has been shown by Rubakov et. al
\cite{rubakov} that a lighter gravitino of the order of a keV could as
well be consistent with relic density observations as a warm dark
matter candidate.  In the presence of bilinear R-parity violation,
gravitino decays into a neutrino and a photon
\cite{yamaguchi,ibarra}.  The decay rate is given by the expression
\cite{takayama} :
\begin{equation}
\label{gravitino-2-body-decay}
\Gamma( \psi_{3/2} \to \gamma \nu) = {\frac{1}{32 \pi}} ~
| U_{\tilde{\gamma} \nu} |^2 ~ {\frac{m_{3/2}^3}{M_P^2}} ,
\end{equation}
where $m_{3/2}$ is the gravitino mass and $M_P$ is the Planck scale,
and $U_{\tilde{\gamma} \nu}$ is given by \cite{ibarra}
\begin{equation}
U_{\tilde{\gamma} \nu} \simeq g_z \left| \sum_{\alpha =1}^4 c_{\tilde{\gamma}\nu} c_{\tilde{z}\alpha}^\star 
{\frac{v_{\nu}}{m_{\chi^0_ \alpha}}} \right| \sim 10^{-8} \left({\frac{\epsilon_3}{10^{-7}}} \right) 
\left(\frac{\tilde{m}}{200 \text{GeV}} \right)^{-1}  \sim 10^{-5} \left({\frac{\epsilon_3}{10^{-4}}} \right) 
\left(\frac{\tilde{m}}{200 \text{GeV}} \right)^{-1},
\end{equation} 
Here, $g_z$ is the gauge coupling, $c_{\tilde{\gamma}\nu}$ and $c_{\tilde{z}\alpha}$ are the mixing elements in the
neutralino-neutrino mixing  matrix, $v_\nu$ is the sneutrino VEV, ${\tilde m}$ is a common mass scale for the superpartners 
and $\epsilon_3$ is the bilinear R-parity
violating parameter defined in Eq.(1). These parameters appearing in the second term of the RHS can be found in
\cite{ibarra}.  Using this formula for the lifetime of a 5 keV gravitino
we find ,
\begin{equation}
\tau^{\text{2-body}}_{3/2} \simeq 3 \times 10^{40} s \left({\frac{\epsilon_3}{10^{-4}}} \right)^{-2} 
\left(\frac{\tilde{m}}{200 ~\text{GeV}} \right)^{2} \left(\frac{m_{3/2}}{5~ \text{keV}} \right)^{-3}. 
\end{equation}
From the above we see that the lifetime of the gravitino is several
orders larger compared to the age of the universe ($\cal{O}$($10^{18}$ s)).

Since, according to the above expressions, their lifetimes are large
but finite, one would expect a few of the axinos/gravitinos to decay
through the two body processes (\ref{axino-2-body-decay}) and
(\ref{gravitino-2-body-decay}).  The resultant photons are in the
X-ray regime and thus could be observed by the astrophysical X-ray
telescopes.  These X-rays would constitute an excess over the
astrophysical background.  In the recent years, there has been a
report of such an observation by the Chandra satellite based
observatory with a line emission at 2.5 $\pm$ 0.11 keV (90\% CL) and
flux of $(3.53 \pm 2.77) \times 10^{-6}$ photons $\text{cm}^{-2}
~s^{-1}$ at (90\% CL) \cite{chandra}.

Recently, Loewenstein and Kusenko \cite{LK} have tried to explain the
Chandra excess \cite{chandra} through a radiative decay of a sterile neutrino of mass
of 5 keV.  It should be noted that astrophysical X-rays have been
observed earlier too from the Suzaku satellite and have led to
stringent constraints on sterile neutrino mixing and mass parameters
\cite{LK2}. In the present work, we would like to obtain similar
constraints on the axino masses and the Peccei-Quinn symmetry breaking
scale $f_{a}$ for a fixed value of the neutralino-neutrino mixing
parameter satisfying neutrino data, using the Chandra
results.  On the other hand, similar analysis on gravitino dark matter
does not provide any constraint because of the larger suppression
through $1/M_{Pl}^2$ in the decay width.

\section{Dark matter halos and flux of X-ray photons}
The X-ray flux from the dark matter halos can be calculated for a
given galaxy knowing its luminosity distance and the dark matter
distribution. In the following we outline the calculation of the flux
in general and apply it for our own galaxy.

An object such as a field galaxy or dwarf galaxy or a cluster of
galaxies possessing a dark matter halo of mass $M_{DM}$ will be
composed of $N = M_{DM}/m_X$ dark matter particles of rest mass
$m_X$. If $\Gamma_X$ is the dark matter particle decay rate into
photons of energy $E_\gamma$, then the total associated X-ray
luminosity is \cite{abazajian-fuller-tucker} 
\bea 
{\cal L} =
\frac{E_\gamma}{m_X}M_{DM}\Gamma_X.  
\eea 
Here we have assumed that the halo is relatively nearby and that
redshift effects on the luminosity are negligible.

We will also consider the simple case where $E_\gamma = m_X/2$ (when
the neutrino mass is negligible compared to that of the dark matter
candidate), which is relevant for the two body decays which are the
focus of the present work.  The flux from an object seen through the
telescope is simply 
\bea 
F =  {{\cal L} \over 4\pi D^2_L, } 
\label{lum}
\eea 
where $D_L$ is the luminosity distance to the object. This can be
taken to be the coordinate distance $r$ from the earth to the dark
matter site (neglecting redshift effect).

With $E_\gamma = m_X/2$ we get
\bea
{\cal L} = \frac{M_{DM}\Gamma_X}{2}. 
\label{luminosity}
\eea
Hence, the flux $F$ can be written as
\bea
F &\approx& {\frac{\Gamma_X}{2}}\int_{l.o.s}{\frac{1}{4 \pi r^2}}\,dM_{DM} \nonumber \\
&=& {\frac{\Gamma_X}{8\pi}}\int_{l.o.s}{\frac{1}{r^2}} \rho_{\rm DM}(r) r^2 \,dr \,d\Omega, 
\label{flux}
\eea 
where $\rho_{\rm DM}(r)$ is the density of dark matter particle as a
function of distance from the galactic centre. In order to calculate
the flux which is coming from the direction of the galactic
anti-center, we have 
\bea 
F = {\frac{\Gamma_X
    \Omega}{8\pi}}\int_{r_\odot}^\infty \rho_{\rm DM}(r)\,dr. 
\label{gac} 
\eea
The flux which is coming from the direction of the galactic center is
given by 
\bea 
F = {\frac{\Gamma_X
    \Omega}{8\pi}}\int_{-\infty}^{r_\odot} \rho_{\rm DM}(r)\,dr. 
\label{gc} 
\eea
Here $r_\odot$ ($\simeq$ 8 kpc) is the distance of an observer on the
earth from the galactic centre (of Milky Way). Note that this
derivation applies to integration along the Milky Way only. The reason
we are considering Milky Way is because of the fact that it is a very
extended source of the dark matter halo and SPI telescope (which has a wide field of view) 
is well suited for such an extended object. Our analysis on SPI telescope will be
presented in Sec. IV and we shall show that this will give a very strong constraint
on the axino dark matter model. 

\begin{figure}
\begin{center}
\includegraphics[height=1.34in,width=1.54in]{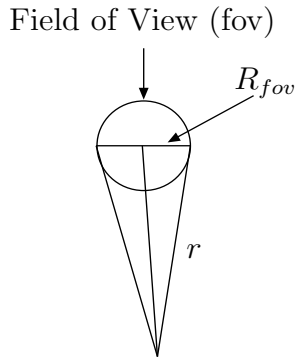}
\vspace*{0.3in}
\caption{Solid angle ($\Omega$) calculation of the dark matter halo.}
\label{solid_angle_calculation}
\end{center}
\end{figure}

In order to determine the dark matter halo contribution to the X-ray
flux, one needs to know the distribution of the DM in the halo of
the Milky Way. In Eq.(\ref{flux}), $\Omega$ is the small ($\ll$ 1)
solid angle corresponding to the field of view of the telescope which
can be calculated from figure \ref{solid_angle_calculation}.
In figure \ref{solid_angle_calculation}, the projected radius of the
field of view is denoted by $R_{fov}$, and the angular field of view
by $\theta$.  Approximating $\Omega$ as the solid angle of a cone with
an apex angle $\theta$, we have 
\beq 
\Omega = {\pi\theta^2\over 4},
\eeq 
where 
\beq 
\theta = { 2 R_{fov} \over r}.  
\eeq 
In particular for ACIS aboard Chandra, $\theta \approx 5 \times 10^{-3}$ rad, which determines
the solid angle. Later on we shall show that a much larger solid angle is obtained for SPI.

The final input needed in the calculation of the flux is
$\rho_{DM}(r)$, the density profile of the dark matter. While the
actual profile of the dark matter in the galaxy is unknown, there are
many profiles put forward in the literature. Many of them also have
analytical expressions.  For example, the isothermal profile
\cite{boyarskyetal} is given by 
\bea 
\rho_{\rm DM}(r) =
{\frac{v^2_h}{4\pi G_N}}{\frac{1}{r^2_c + r^2}}, 
\eea 
where $v_h$ corresponds to the contribution of the DM halo into the
galactic rotation curve in its flat part, $v_h \approx$ 170 km/s. Here
the core radius $r_c$ is almost 4 kpc. Now, the profile integral in
Eq.(\ref{gac}) for the galactic anti-centre is 
\bea
\int_{r_\odot}^\infty \rho_{\rm DM}(r)\,dr = {\frac{C}{r_c}}
\left[\frac{\pi}{2} - \tan^{-1}\left(\frac{r_\odot}{r_c}\right)
\right] =
{\frac{C}{r_c}}\tan^{-1}\left(\frac{r_c}{r_\odot}\right),
\label{flux-anticenter}
\eea
where $C = {v^2_h/4\pi G_N}$. Similarly for the galactic centre case
from Eq.(\ref{gc}):
\bea 
\int_{-\infty}^{r_\odot} \rho_{\rm
  DM}(r)\,dr &=& {\frac{C}{r_c}} \left[\frac{\pi}{2} +
  \tan^{-1}\left(\frac{r_\odot}{r_c}\right) \right].
\label{flux-center}
\eea 
Using Eqs.(\ref{flux-anticenter}) and (\ref{flux-center}) one can
calculate the flux from the galactic anti-center and galactic center,
respectively.

Once again let us mention that the calculation of the dark matter
line flux in the direction of the Galactic center or anti-center
has been performed just for simplicity. In general one can calculate
the flux coming from any given direction on the sky. We shall present
our analysis for an angular distance 13$^\circ$ between the given direction on
the sky and the direction towards the Galactic center when we present our results
for SPI. 

Another point is that in Eq.(12) and (13) the upper (lower) limit of integration 
is actually not infinte but depends on the assumed size of the Milky Way dark matter halo.
However, since the integral is converging, we shall obtain the same result as long as the
limit is 200 kpc or above. This is the reason we have assumed the upper (lower) limit to be 
infinte just for simplicity.

\begin{figure}
\begin{center}
\includegraphics[width=0.8\textwidth]{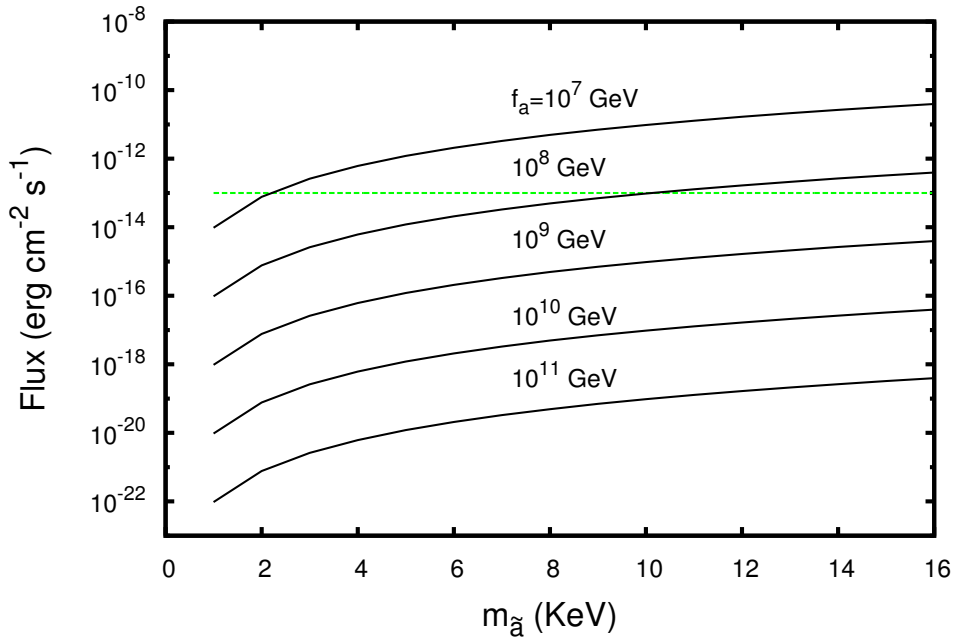}
\caption{Flux of photons from the Milky Way galactic centre for the
  Max. disk model $A_2$. The horizontal line is the lower limit of
  detectability of the flux applicable for the Chandra X-ray satellite
  observations as explained in the text.}
\label{flux_photon_mw_A2_centre}
\end{center}
\end{figure}

The data from Virgo cluster by ACIS aboard Chandra has been analyzed in Ref. \cite{abazajian-fuller-tucker}.
The instrumental background has been estimated to be $2 \times 10^{-2}$ 
cts $s^{-1}$ in a 200 eV energy bin. It has been shown that in order to detect a dark matter decay line
signal of energy $E_\gamma = m_X/2$ at 4$\sigma$ level, one must observe a flux
above the value, $F_{det} = 10^{-13}~{\rm erg}~{\rm cm}^{-2}{\rm sec}^{-1}$
with an integration time of 36 ksec for observation. In our analysis of the 
Milky Way dark matter halo, we shall assume that the same flux limit is applicable
in the entire x-ray energy range of 1 keV to 8 keV relevant for Chandra telescope.  
This determines the region of the parameter space for any theoretical scenario, which 
can be probed by Chandra.

\section{Numerical Analysis for Chandra Telescope}

We have taken the value of the neutrino-neutralino mixing parameter
($\xi_i U_{{\tilde \gamma}{\tilde Z}} \equiv U_{{\tilde \gamma}\nu}$)
to be $9 \times 10^{-7}$, which is the allowed value from neutrino
data.  In figure \ref{flux_photon_mw_A2_centre}, we show the flux from
the Milky Way galactic centre as a function of the axino mass ${\rm
  m}_{\tilde{\rm a}},$ following the Max. disk model $A_2$
\cite{klypin2002} of NFW profile \cite{nfw1,nfw2} for various values
of $f_a$.  As can be seen from the figure, the lower is the values of
$f_{a}$, the larger is the flux. From the figure we see that for the
astrophysically unconstrained range of the Peccei-Quinn symmetry
breaking scale, namely, $f_{a} \gsim 10^9$ GeV
\cite{astrophysical_constraints}, the axino mass is unrestricted in
the range shown in the figure. However, for lower values of $f_a$ some
constraint can be obtained.

\begin{figure}
\begin{center}
\begin{tabular}{cc}
\includegraphics[width=0.5\textwidth]{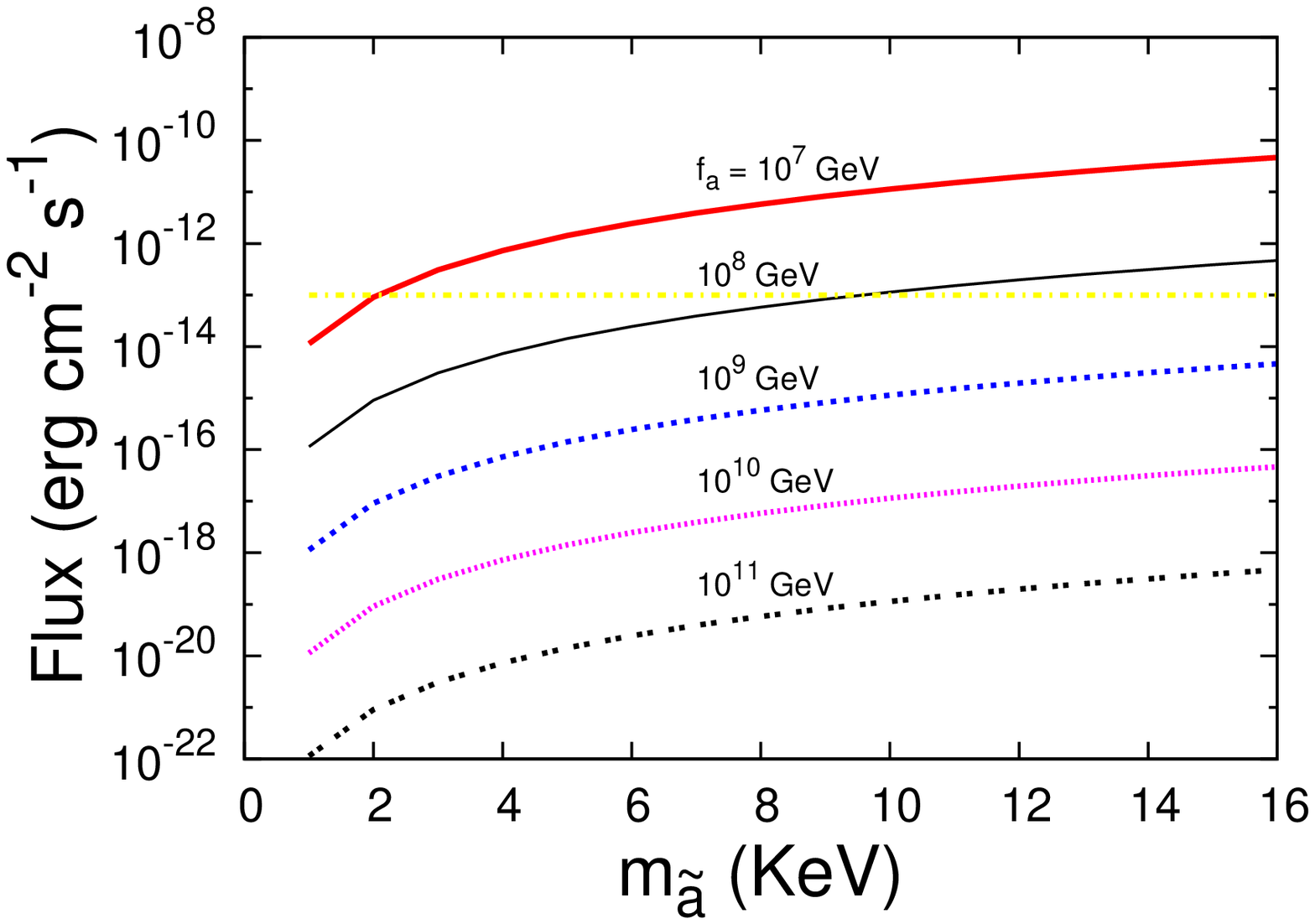}
\includegraphics[width=0.5\textwidth]{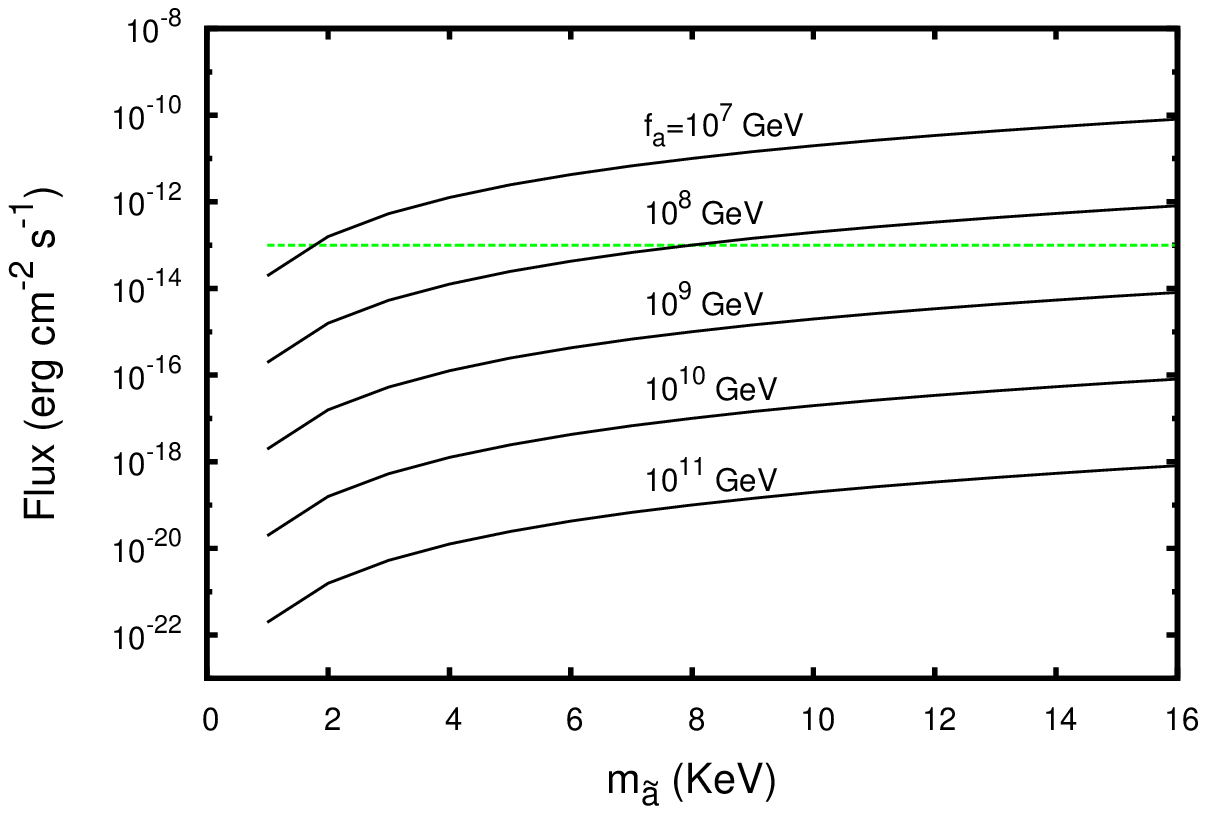}\\ \nonumber 
\includegraphics[width=0.5\textwidth]{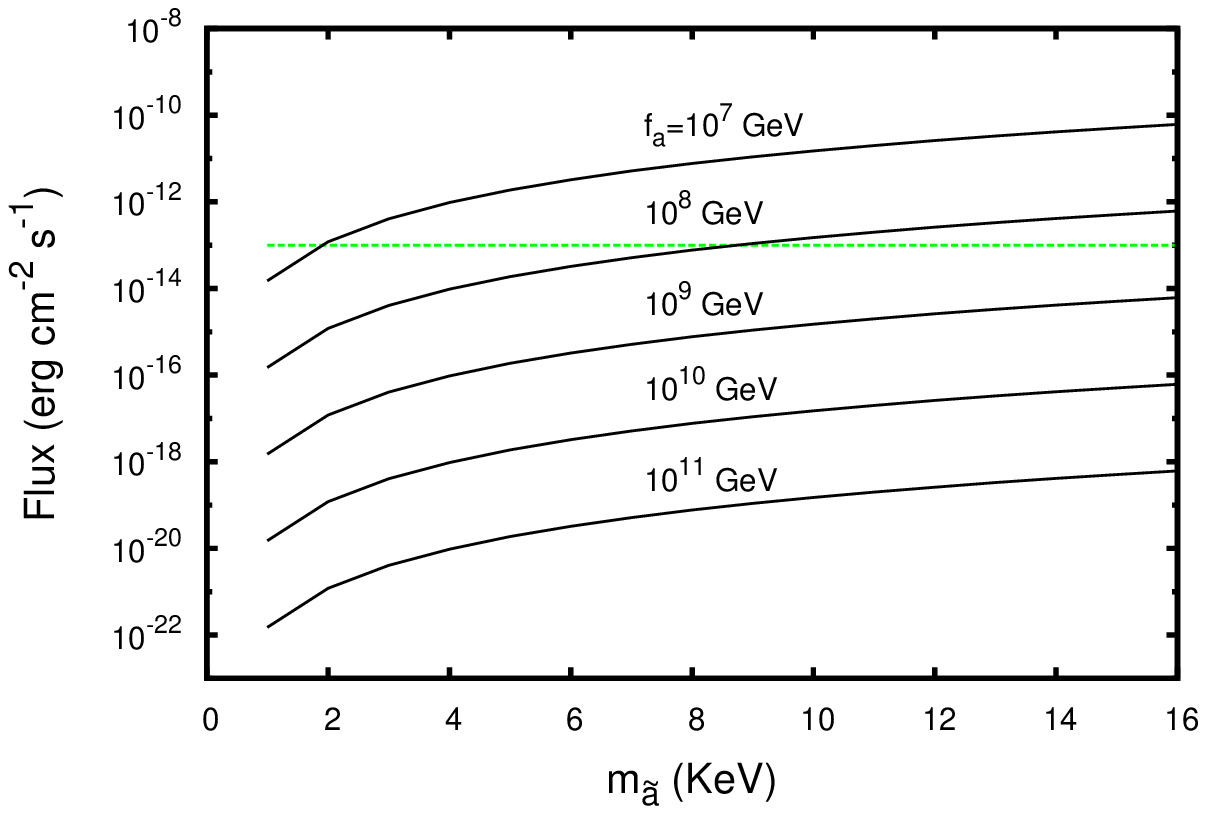} 
\includegraphics[width=0.5\textwidth]{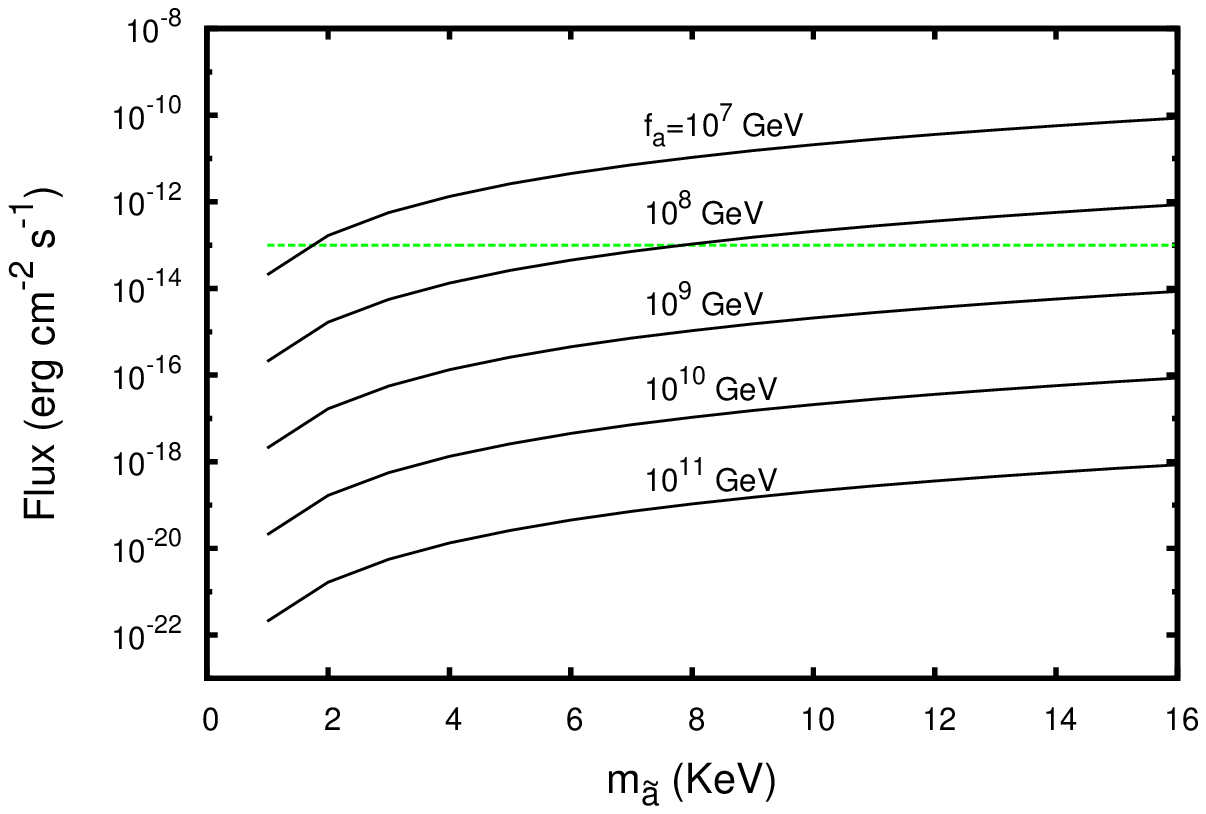} 
\end{tabular}
%\vspace*{-1.0in}
\caption{Flux of photons from the Milky Way galactic centre for
  various profiles. Top left is for isothermal profile, top right is
  for the NFW favored models ($A_1$ or $B_1$) \cite{klypin2002}. Similarly, bottom left is for the 
  NFW Max. disk models $B_2$ of the MW DM halo whereas bottom right is for the NFW
  profile best fit parameters given by Battaglia et al \cite{battaglia}. As before the horizontal line 
  is the lower limit of the flux applicable for the Chandra X-ray satellite observations.}
\label{flux_photon_mw_various_centre}
\end{center}
\end{figure}

The question that arises is$\colon$ for which profiles there exist
constraints from Chandra within the astrophysically unconstrained
values of $f_{a}$. In figure \ref{flux_photon_mw_various_centre} we
present the fluxes for various profiles including the isothermal
profile discussed in the previous section. The isothermal
\cite{boyarskyetal}, and the NFW profiles \cite{klypin2002,battaglia}
do not provide constraint if one takes the lower limit on
$f_{\tilde{a}} \gtrsim 10^{9}$ GeV strictly. However, there are ways
of avoiding these astrophysical constraints in some situations such as
(a) existence of paraphotons
\cite{masso-redondo-paraphoton,masso-redondo-prl} and (b) existence of
temperature dependent couplings
\cite{environment-dependent-coupling}. In either case, $f_a$ lower
limit can be relaxed, sometimes even to values of
$\mathcal{O}(10^{6})$ GeV. In such situations the constraints from
Chandra could play an important role.

\begin{figure}
\begin{center}
\includegraphics[height=3.54in,width=5.54in]{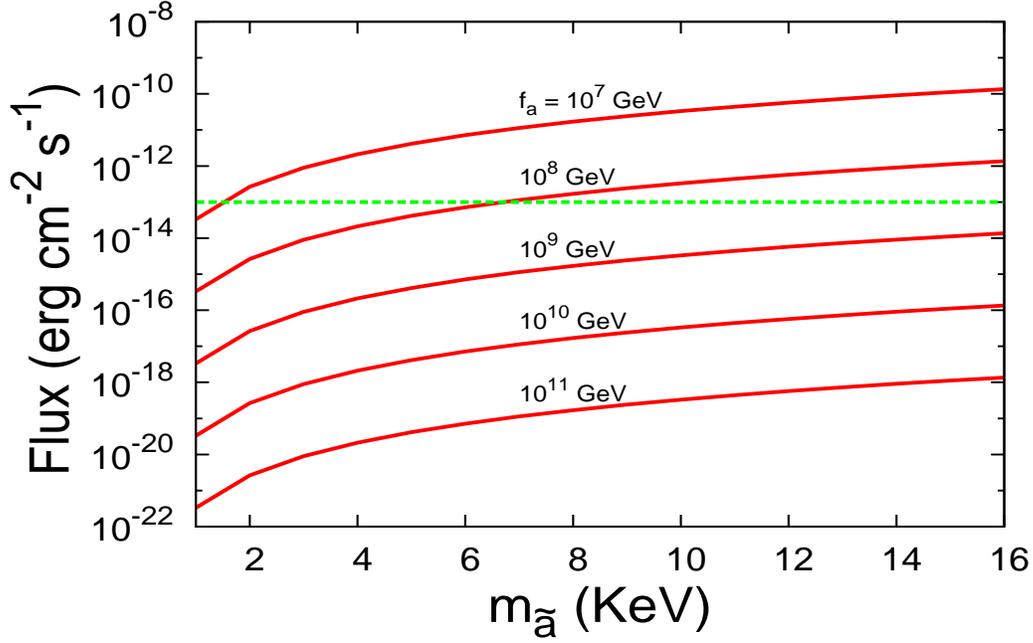}
%\vspace*{-1.0in}
\caption{Flux of photons from the centre of the Virgo cluster.  As
  before the horizontal line is the lower limit of the flux applicable
  from the Chandra X-ray satellite observations.}
\label{distant_galaxy}
\end{center}
\end{figure}

Before concluding let us briefly touch the constraints on our model
coming from the observation of the Virgo clusters 
\cite{abazajian-fuller-tucker,boyarskyetal-prd,galaxy-cluster-refs}.
Using Eqs.(\ref{axino-life-time}), (\ref{lum}) and (\ref{luminosity}),
the flux from a cluster can be written as,
\bea 
F \approx
2.19 \times 10^{-21}~{\rm ergs}~{\rm cm}^{-2}{\rm s}^{-1}
\left(\frac{D_L}{1~{\rm Mpc}} \right)^{-2} \left(\frac{M^{\rm
      fov}_{\rm DM}}{10^{11} M_\odot} \right) \left(\frac{U_{{\tilde
        \gamma}\nu}}{10^{-7}} \right)^2 \left(\frac{m_{\tilde a}}{5
    {\rm keV}} \right)^3 \left(\frac{f_a}{10^{11}} \right)^{-2}.
\label{distant_galaxy_flux}
\eea 
The values for $D_L$ and $M^{\rm fov}_{\rm DM}$ for the Virgo cluster
are given in \cite{abazajian-fuller-tucker}. We have used $M^{\rm
  fov}_{\rm DM} = 10^{13}M_{\odot}$ and $D_L = 20.7$ Mpc in
Eq.(\ref{distant_galaxy_flux}) and demanded that the flux should be
less than the detectable limit $10^{-13}$ ergs cm$^{-2}$ s$^{-1}$. The
results are plotted in figure \ref{distant_galaxy}. It is evident that
the constraints obtained, are at least as stringent as those for the
Milky Way, for the more optimistic density profiles.  Since the
luminosity distance $D_L$ is much larger for the Virgo cluster, one
would expect that the corresponding flux will be smaller than that
obtained for Milky Way. For the Virgo cluster, however, the total dark
matter mass in the field of view ($M^{\rm fov}_{\rm DM}$) is much
larger compared to that of the Milky Way in a particular direction
(say, the direction of the galactic center or anti-center). This large
mass for Virgo cluster compensates for the large luminosity distance
and hence the resulting flux becomes comparable to the flux from Milky
Way for more optimistic density profiles such as the NFW favored model
$A_1$ or $B_1$ \cite{klypin2002}.

\section{Numerical analysis for SPI}

In this section we shall look at the bounds on the axino parameters
originating from an analysis based on the data from the
high-resolution spectrometer SPI on board the INTEGRAL sattelite
\cite{oleg-spi}. SPI is sensitive to X-ray photons in the energy range
20 keV to 7 MeV. Obviously, the data from SPI is applicable to axinos
in the mass range of 40 keV to 14 MeV.  However, we shall restrict
ourselves to the axino mass range of 40 keV to 400 keV, which includes
the window on the axino mass where the axino can act as a warm dark
matter candidate. The negative search of a dark matter decay line
using SPI put restrictions on the dark matter decay line flux
\cite{oleg-spi}. Using this flux restrictions we derive bound on the
axino mass parameters.

In our simple minded analysis, we have used the {\it partially coded field of view}
of the SPI detector given by $\theta_{PCFOV} \approx 35^\circ$. This corresponds to
a solid angle $\Omega = 2 \pi (1 - \cos(\theta_{PCFOV}/2)) \approx 0.29$ and we have used
this value of the solid angle for our numerical calculation. The dark matter decay line flux
has been calculated using various profiles mentioned in the previous section. In addition,
we have used the {\it minimal model} described in Ref.\cite{oleg-spi}, for which the
dark matter density is constant within $r_\odot$ whereas it is given by the model $A_2$
of Klypin et al in Ref. \cite{klypin2002} for $r > r_\odot$.

The results obtained in Ref.\cite{oleg-spi} using the SPI data set is
applicable to any model of decaying dark matter. We have used the
results shown in figure 11 of Ref.\cite{oleg-spi}, for the 3$\sigma$
restrictions on the lifetime of the dark matter particle as a function
of the X-ray photon energy \footnote{We sincerely thank Oleg
  Ruchayskiy for providing us with the data file corresponding to
  figure 11 of their paper.}.  Using the {\it minimal model} for the
dark matter profile as mentioned above, we translate this restriction
on the lifetime into a restriction on the decay line flux as a
function of the axino mass. The dark matter decay line flux is also a
function of the angular distance $\phi$ between the direction along
the line of sight and the direction towards the Galactic center. We
have used $\phi = 13^\circ$ (the inner Galaxy) for the calculation of
the flux restriction.

\begin{figure}
\begin{center}
\includegraphics[width = 0.8\textwidth]{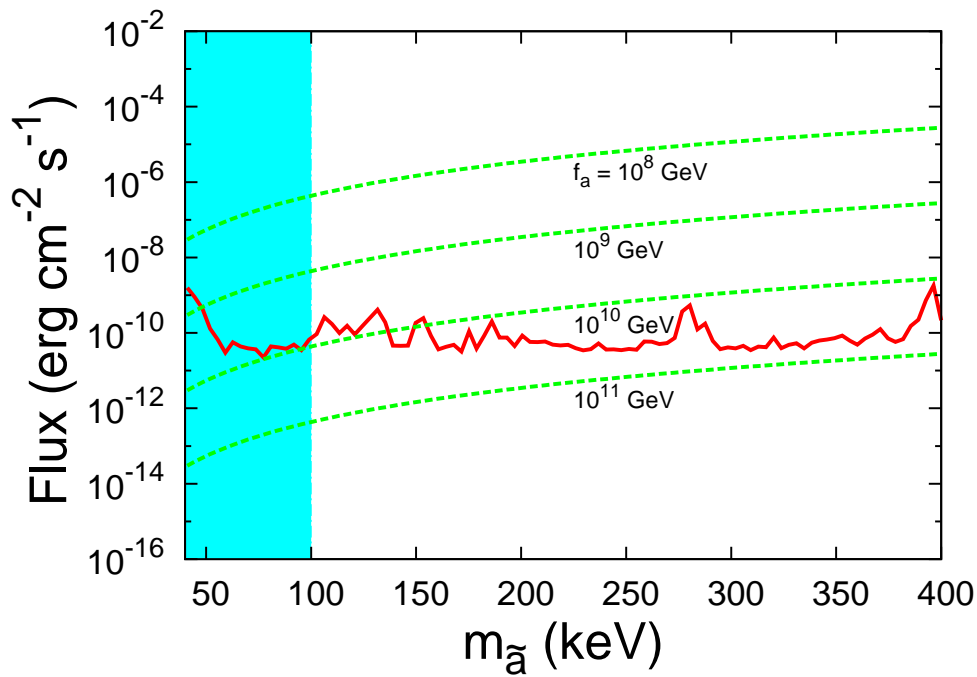}
\caption{Flux of photons from the Milky Way inner Galaxy ($\phi = 13^\circ$)
  for the minimal profile described in the text.  The red zigzag line is 
  the lower limit of the flux coming from the negative search of a dark matter 
  decay line using the data from SPI spectrometer on board the
  INTEGRAL satellite. The shaded region is the window on the axino
  mass where the axino can act as a warm dark matter candidate.}
\label{flux_photon_mw_minimal_profile}
\end{center}
\end{figure}

In figure \ref{flux_photon_mw_minimal_profile} we show the flux restriction using the red zigzag curve. In addition, 
in the same figure we show the line flux as a function of the axino mass coming from an axino dark matter, for different 
values of $f_a$. These curves are drawn for the minimal profile of \cite{oleg-spi}. 

From the figure we see that for the astrophysically unconstrained range of the Peccei-Quinn
symmetry breaking scale, namely, $f_a \gsim 10^9$ GeV \cite{astrophysical_constraints}, 
axino mass $m_{\tilde a} \gsim$ 50 keV could be excluded by SPI data.

\begin{figure}
\begin{center}
\includegraphics[width = 0.8\textwidth]{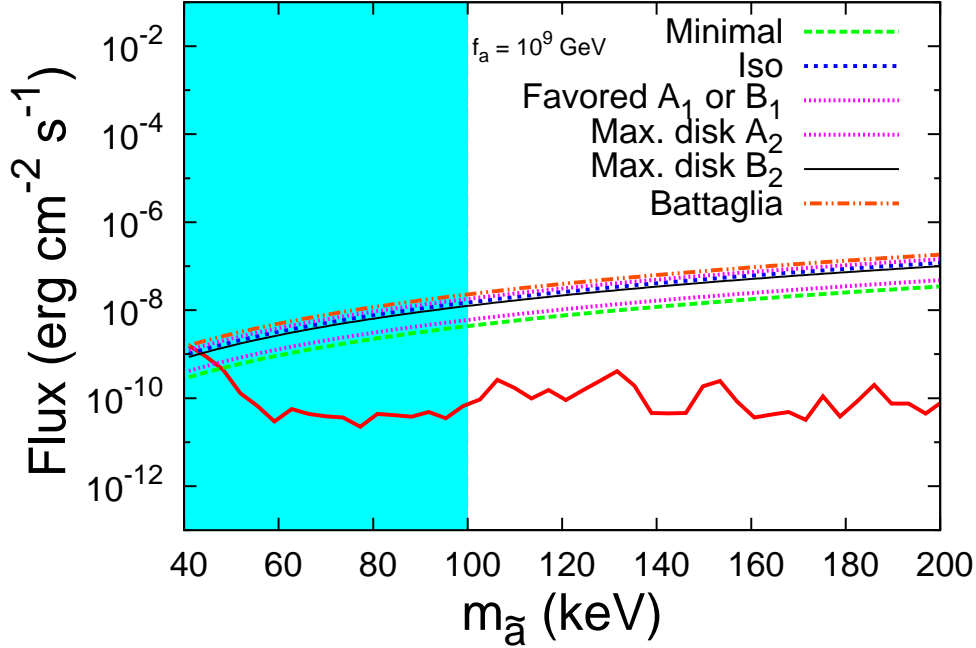}
\caption{Flux of photons from the Milky Way inner Galaxy ($\phi = 13^\circ$) for
  all profiles for $f_{a} = 10^{9}$ GeV.  As before, the red zigzag
  line is the lower limit of the flux coming from the SPI
  spectrometer observations. The shaded region is the window on the axino
  mass where the axino can act as an warm dark matter candidate.}
\label{flux_photon_mw_allprofile_spi}
\end{center}
\end{figure}

On the other hand, the results depend on the profile one chooses for the dark matter in the
Milky Way. In particular, this dependence is more pronounced in the direction of the
inner Galaxy. In figure \ref{flux_photon_mw_allprofile_spi} we plot the flux for a fixed value of $f_a$ 
for the isothermal profile and different models of the NFW profile (the values of the different 
parameters appearing in these profiles are taken from \cite{oleg-spi}) as well as the minimal profile of
\cite{oleg-spi}. The shaded region and the zigzag line have the same meaning as before. From the figure 
it is obvious that there could be some noticeable variation in the flux depending on the profile one uses. 
Accordingly this gives us different upper bounds on the axino mass parameter.
However, this variation becomes more and more negligible for some of the profiles if one looks at the direction 
of the Galactic anti-center.  

\section{Summary, conclusions and outlook}

We have attempted to constrain the parameter space of a warm dark
matter candidate from the non-observation of galactic X-ray line flux
at the X-ray telescopes such as Chandra and SPI. We are considering an R-parity violating
SUSY scenario, where the gravitino and the axino are the two likely
candidates of warm dark matter.  Due to the violation of lepton
number, they can have {\it some} two body decays into a
(monochromatic) photon and a neutrino, although their mean lifetimes
are larger than the age of the universe. These photons correspond to
X-ray lines that are potentially observable at X-ray telescopes.

We have found that gravitino decays in the above channel yield too
small a flux to lead to any useful constraint, due to the suppression
of gravitino interaction by the Planck mass. The axino, on the other
hand, couples with a strength inversely proportional to the
Peccei-Quinn symmetry breaking scale ($f_a$) , which can lie in the
range $10^{9 - 12}$ GeV, and is therefore more promising from the
viewpoint of X-ray satellite observations. Our analysis is based on the
assumption that R-parity violation, triggered by bilinear terms in the
superpotential, is of such a degree as to accommodate the pattern of
neutrino masses and mixing suggested by the solar and atmospheric
neutrino data. On this basis, we find that the limits from Chandra and SPI
disallow a sizable part of the yet allowed axino parameter space
defined by the axino mass ($m_{\tilde a}$) and $f_a$. In particular,
axino masses in the range $\sim$ 50 - 100 keV tend to be disallowed by SPI. The best
results follow from a study of X-rays from the inner part of the Milky
Way. For galaxy clusters such as the Virgo cluster a similar
analysis has been performed for Chandra with similar conclusions as in the case 
of Milky Way. 

If one uses the exact expressions for neutrino masses, there could be
some SUSY parameter space dependence of the above results. In
particular, there could be some cancellations in the expression for
sneutrino VEV in some regions of the parameter space. However, 
the general conclusions, based on the limits on the extent of
R-parity violation, are unlikely to differ drastically from
what have been derived above.

It should be noted that the limits have some dependence on the profile
of the dark matter. The flux from the inner part of the galaxy can vary 
significantly depending on the profile. For some profiles like
NFW favored models $A_1$ or $B_1$, NFW parameters of Battaglia et al, there could be constraints for $f_a
= 10^{9}$ GeV and ${\rm m}_{\tilde{\rm a}} \gtrsim 42$ keV. Even constraints can be obtained
for $f_a = 10^{10}$ GeV from SPI data.
On the whole, for the
lower range of allowed $f_a$, namely, $10^{9 - 10}$ GeV, axino masses
above 42-60 keV (depending on $f_a$) can be ruled out from the
SPI data on Milky Way.
 
At the end, we would like to remark that the Chandra observatory is a
high resolution experiment, but its energy range for X-rays is
restricted to 8 keV. Similarly, for SPI the energy range is 20 keV to 7 MeV. 
There are other experiments; among them,  XMM-Newton 
has less resolution but energy range up to 12 keV
\cite{nasa-website-xmm}. Beppo-SAXX can detect X-rays up to 15 keV
\cite{bepposax}. Old experiments like HEAO-I, and HEAO-III can measure
X-rays up to 10 MeV. HEAO-II (Einstein) can measure X-rays up to 20
keV \cite{nasa-website}. Japan's fifth X-ray satellite Suzaku can carry out 
high resolution spectroscopic observations in a wide energy range of 0.3 keV to 600 keV 
\cite{suzaku}. A global analysis based on all of these
experiments can lead to further insight into decaying warm dark matter
candidates in an R-parity violating scenario. 

%-------------------------------------------------------------------------
\acknowledgments{We thank Oleg Ruchayskiy for many helpful discussions. 
The work of BM was partially supported by funding
  available from the Department of Atomic Energy, Government of India,
  for the Regional Centre for Accelerator-based Particle Physics
  (RECAPP), Harish-Chandra Research Institute. SKV acknowledges
  support from DST project ``Complementarity between direct and
  indirect searches for Supersymmetry''. PD is supported through the
  Gottfried Wilhelm Leibniz Program by the Deutsche
  Forschungsgemeinschaft (DFG). PD, BM and SR thank Centre for High
  Energy Physics, Indian Institute of Science, Bangalore, for
  hospitality during the initial stages of this work. SKV and BM
  acknowledge the hospitality of the Department of Theoretical
  Physics, Indian Association for the Cultivation of Science, Kolkata,
  while this work was on. SR and SKV also thank RECAPP for hospitality
  during the final stages of this work. SR thanks the CERN Theory group
  for hospitality during the preparation of the revised version of the work.}
%%%%%%%%%%%%%%%%%%%%%%%

%%%%%%%%%%%%%%%%%%%%%%%%

\end{document}